\documentclass[12pt,prb,onecolumn]{revtex4}
\usepackage{graphicx,amsmath,amssymb,times}
\begin{document}

\renewcommand{\Re}{\mathop{\mathrm{Re}}}
\renewcommand{\Im}{\mathop{\mathrm{Im}}}
\renewcommand{\b}[1]{\mathbf{#1}}
\renewcommand{\c}[1]{\mathcal{#1}}
\renewcommand{\u}{\uparrow}
\renewcommand{\d}{\downarrow}
\newcommand{\bsigma}{\boldsymbol{\sigma}}
\newcommand{\blambda}{\boldsymbol{\lambda}}
\newcommand{\tr}{\mathop{\mathrm{tr}}}
\newcommand{\sgn}{\mathop{\mathrm{sgn}}}
\newcommand{\sech}{\mathop{\mathrm{sech}}}
\newcommand{\diag}{\mathop{\mathrm{diag}}}
\newcommand{\half}{{\textstyle\frac{1}{2}}}
\newcommand{\sh}{{\textstyle{\frac{1}{2}}}}
\newcommand{\ish}{{\textstyle{\frac{i}{2}}}}
\newcommand{\thf}{{\textstyle{\frac{3}{2}}}}

\bibliographystyle{naturemag}

\title{Spin polarization of the quantum spin Hall edge states}

\author{Christoph Br\"{u}ne$^1$, Andreas Roth$^1$, Hartmut Buhmann$^1$, Ewelina M. Hankiewicz$^2$, Laurens W. Molenkamp$^1$, Joseph Maciejko$^{3,4}$, Xiao-Liang Qi$^{3,4,5}$ and Shou-Cheng Zhang$^{3,4}$}

\affiliation{$^1$Physikalisches Institut (EP3) and R\"{o}ntgen
Center for Complex Material Systems, Universit\"{a}t W\"{u}rzburg,
Am Hubland, 97074 W\"{u}rzburg, Germany\\
$^2$ Institut f\"{u}r Theoretische und Astrophysik and R\"{o}ntgen Center for
    Complex Material Systems, Universit\"{a}t W\"{u}rzburg,
    Am Hubland, 97074 W\"{u}rzburg, Germany\\
$^3$Department of Physics, Stanford
University, Stanford, CA 94305, USA\\
$^4$Stanford Institute for Materials and Energy Sciences,\\
SLAC National Accelerator Laboratory, Menlo Park, CA 94025, USA\\
$^5$Microsoft Research Station Q, Elings Hall, University of California, Santa Barbara, CA 93106, USA}

\begin{abstract}
While the helical character of the edge channels responsible for charge transport  in the quantum spin Hall regime of a two-dimensional topological insulator is by now well established, an experimental confirmation that the transport in the edge channels is spin-polarized is still outstanding. We report experiments on nanostructures fabricated from HgTe quantum wells with an inverted band structure, in which a split gate technique allows us to combine both quantum spin Hall and metallic spin Hall transport in a single device. In these devices, the quantum spin Hall effect can be used as a spin current injector and detector for the
metallic spin Hall effect, and vice versa, allowing for an all-electrical detection of spin polarization.
\end{abstract}

\maketitle
\section{Introduction}

The discovery that HgTe quantum wells (QWs) with an inverted band structure are 2-dimensional topological insulators has generated a great interest in this novel state of quantum matter\cite{bernevig2006d,konig2007,Konig2008}. When the
thickness $d$ of the HgTe QW is increased beyond a critical value $d_{c}$, a quantum
phase transition turns a conventional insulator into its topologically non-trivial counterpart.
In this so-called quantum spin Hall (QSH) phase\cite{kane2005A,bernevig2006a}, current-carrying states are confined at the edge of the sample, while the bulk is insulating. These edge states are protected against backscattering from non-magnetic impurities\cite{wu2006,xu2006,Maciejko2009} and their propagation direction is helical, i.e. that opposite spin states counter-propagate along a given edge of the sample.
 When the applied gate
voltage places the Fermi level inside the bulk gap, two-terminal transport
experiments measure a quantized conductance of $2e^2/h$ with $e$
the electron charge and $h$ the Planck constant, independent of
the sample width, which constitutes strong evidence for edge state
conduction\cite{konig2007}. More recent nonlocal transport
measurements in the QSH regime unambiguously establish that
transport occurs through topologically protected edge
channels\cite{roth2009,buttiker2009}. While edge state
conduction in the QSH regime is thus experimentally well established, there exists
so far no direct experimental evidence that the transport in the helical edge states
of 2D topological insulators is spin-polarized, which is a
fundamental characteristic of this new state of matter.

In this work, we construct novel devices (Figs.~1 and 2a) that
enable us to study the spin polarization of the QSH edge states by
purely electrical means. First
of all, these devices allow us to detect the spin polarization of
the QSH edge states (Fig.~1b) via the inverse spin Hall
effect\cite{hankiewicz2004,valenzuela2006,Bruene2010} (SHE$^{-1}$).
Second, our devices enable us to show that because of their
helical nature, the QSH edge states can be used as a detector of
spin current (Fig.~1a). In our devices, the spin current is
generated by the intrinsic ballistic spin Hall
effect\cite{hankiewicz2004,Rothe2010} (SHE)
exhibited by a HgTe QW in the metallic regime \cite{Bruene2010}.
These two experiments establish for the first time
spin polarization of the helical edge states in topological insulators,
and also demonstrate potential applications of the QSH effect for spintronic devices.

\section{Principle of the experiment}
\begin{figure}[t]
\includegraphics[width=4.5in]{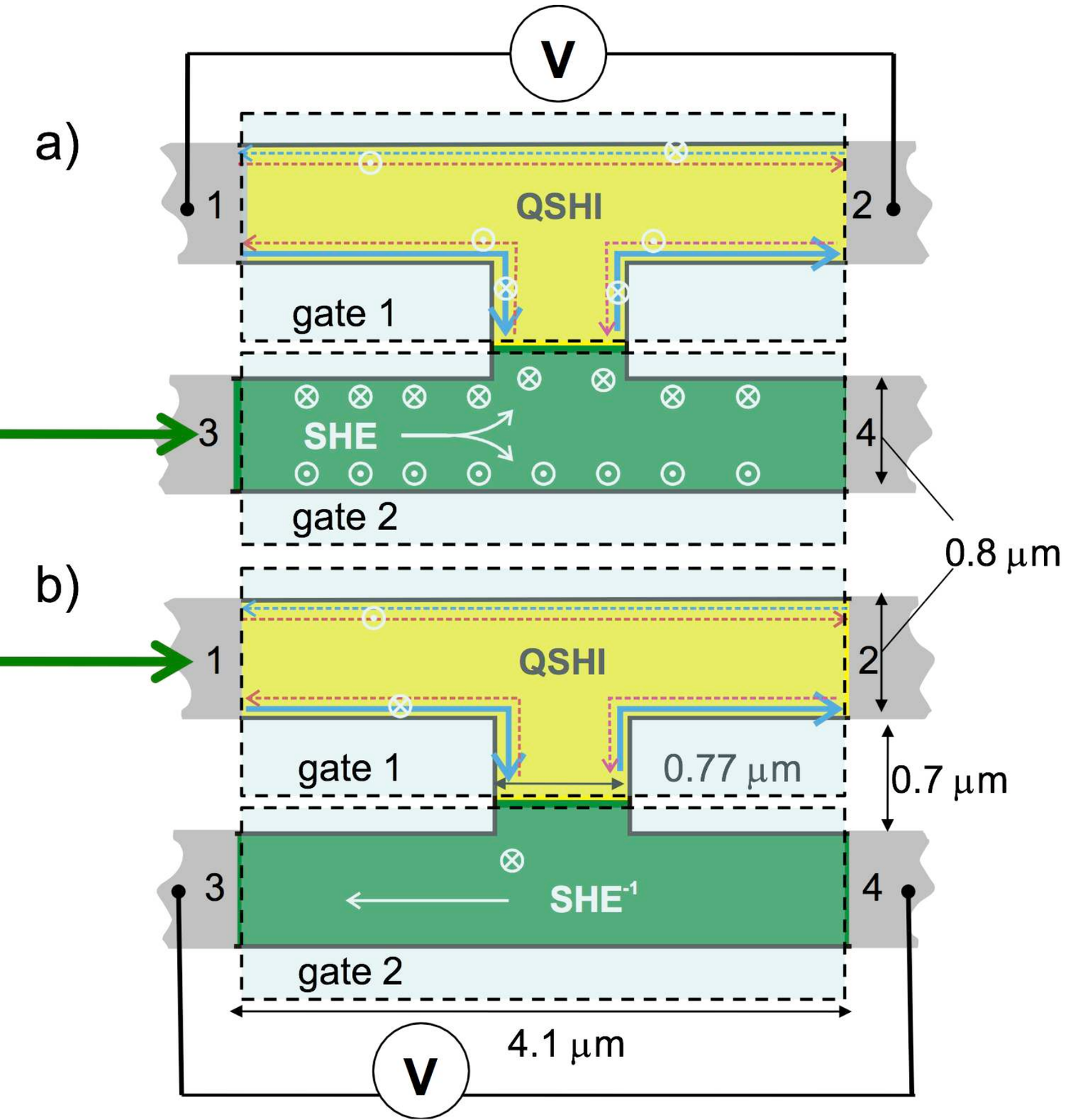}
\caption{Schematic layout of the two experiments on split-gated H-bar devices discussed in this paper.
The dimensions indicated are those of the actual devices used in the experiments.
a) shows the configuration where the current is injected
into a metallic region (green, contacts 3 and 4). The spin-orbit interaction leads, through the
spin Hall effect (SHE), to spin accumulation at the edges of the leg, as indicated schematically. The
upper part of the structure (yellow) is in the QSH regime; the difference in chemical potential between the two spin states
in the interface region is transferred by the helical edge channels to voltage
contacts 1 and 2.
In b) the injector and detector regions are interchanged: the current is
injected into the (spin-polarized) helical edge channels of the upper leg, causing a spin accumulation
in the lower metallic region. The inverse spin Hall effect (SHE$^{-1}$) converts the spin accumulation into
a voltage signal.}
\end{figure}
Before presenting our results, we first describe the principle of
our experiment in more detail. Since the magnetic field
originating from spin polarized carriers in helical edge channels
is too small to be detected directly, we have designed an
experiment that converts magnetic information into an electrical
signal. Figure~1 illustrates the idea of the experiments, which
are performed on an H-shaped mesa structure (which we call
`H-bar') in which the carrier concentration in the two legs of the
`H' can be adjusted separately. Consider the situation illustrated
in Fig.~1, where the bottom leg is metallic (indicated by the
green color, and either $n$- or $p$-type) and the top leg is tuned
into the QSH regime (indicated by the yellow color), with
the counter-propagating helical edge channels depicted as blue and
red trajectories. We perform two separate complementary
experiments.

In Fig.~1a, the current is injected into the metallic part of the
structure (contacts 3 and 4) while a voltage signal is detected
across the top leg (contacts 1 and 2), which is gated into the QSH
insulator state. The inverted band structure in HgTe results in a
large spin-orbit coupling\cite{novik2005,Rothe2010}, which has
previously enabled us to observe a ballistic intrinsic SHE in a
small H-bar structure with a homogeneous carrier
profile\cite{Bruene2010}. Similarly, when in the experiment of
Fig.~1a a charge current is injected into the metallic leg, the
intrinsic SHE will induce a separation of carriers with opposite
spin polarizations toward opposite edges of this leg. This leads
to a difference in chemical potential for opposite spin states in
the area where the metallic part of the structures borders on the
QSH region. The spin polarized helical edge channels coming from
the QSH region couple selectively to the chemical potential
of the matching spin species in the metallic region and transfer
this difference in potential to voltage contacts 1 and 2. For
non-spin-selective edge channels the voltage signal is expected to
be zero, while for the spin polarized QSH edge channels we expect
a nonzero signal. Thus the observation of a nonlocal signal in
this configuration is evidence that the metallic leg develops an
intrinsic SHE, as well as that the helical edge channels are spin
polarized in the QSH insulator regime.

In the reverse configuration of Fig.~1b, the current is injected
(contacts 1 and 2) into the area of the sample that is gated into
the QSH regime, while a nonlocal voltage drop is measured across
the metallic leg (contacts 3 and 4).
In this configuration, the spin polarized helical edge channels
inject a spin polarized current into the metallic leg, causing a
local imbalance in the chemical potential of spin-up and spin-down
polarized carriers. Due to the SHE$^{-1}$ (see
Refs.~\cite{hankiewicz2004,valenzuela2006,Bruene2010}),
the spin current in the metallic region induces
a voltage between contacts 3 and 4. Again, this voltage can only
develop provided the helical edge channels are spin polarized, and
the metallic leg exhibits the SHE$^{-1}$.

A possible complication in both of the above experiments is the
detection of a stray spreading voltage. In the configuration of
Fig.~1a, this could result from a voltage drop in the metallic leg
along the interface with the area in the QSH insulator regime,
while in Fig.~1b, the finite distance between in- and outgoing
edge channels at this interface could produce a similar effect.
However, in practice such stray voltages are strongly reduced by
the exact layout of the experiment, the quasi-ballistic nature of
the transport in the metallic leg and the finite width of the edge
channels. In the supplementary online material (Section VI), we
present theoretical models and experimental evidence showing that
any residual stray voltages can be neglected as compared with the
actual signals obtained in the experiments described below.

\section{Sample fabrication and transport characterization measurements}

Our H-bar structures are fabricated from inverted HgTe/HgCdTe
type-III QWs with a nominal well width of 9~nm, located 74~nm
below the surface. At  a temperature of 4~K (and for a grounded gate),
the carrier density is $n\approx4\times10^{11}$~cm$^{-2}$ . The
carrier mobility is then
$\mu\approx1.1\times10^5$~cm$^2/$(V$\cdot$s), yielding an elastic
mean free path larger than 2~$\mu$m. The devices are patterned
using optical and electron beam lithography, with dimensions as
indicated in Fig.~1. In order to control the carrier density, the
device is gated by Au gate electrodes separated from the sample
surface by a 110~nm thick insulating Si$_3$N$_4$/SiO$_2$
multi-layer stack. By applying a voltage $V_\mathrm{gate}$ to the
top gates, the electron carrier density of the QW can be adjusted,
going from an $n$-type behavior for $V_\mathrm{gate}>0$ through
the bulk insulator state into a $p$-type regime for
$V_\mathrm{gate}<0$. For reasons of comparison, the experimental
data in Fig.~2, 3 and 4 are plotted as a function of a normalized
gate voltage $V^{*}_\mathrm{gate}\equiv V_
\mathrm{gate}-V_\mathrm{thr}$, where the threshold voltage
$V_\mathrm{thr}$ is defined as the voltage for which the
resistance is largest in a particular fixed reference measurement.
As is evident from the characterization data in Fig.~2b and c,
which were obtained from a Hall bar fabricated from the same wafer
material as the H-bar nanostructures, we find that for gate
voltages
 $V^{*}_ \mathrm{gate} \gtrsim 0.5$~V the QW is $n$-type metallic, and for $V_ \mathrm{gate} \lesssim -0.5$~V
 it is $p$-type metallic. The split-gate design (gates 1 and 2) of Fig.~1 provides an
independent control of the carrier density for each leg of the
H-bar structure, enabling us to gate one part of the sample into
the QSH insulator regime and the other part into either
 $n$- or $p$-type metallic regimes. An electron microscope picture
of the actual device structure is shown in Fig.~2a. The transport
measurements are done at a constant temperature of 1.8~K employing
quasi-dc low frequency (13~Hz) lock-in techniques using a voltage
bias below 100~$\mu$V.

\section{Experimental results}

While experiments have been performed on a variety of different
devices and yield similar results, for reasons of consistency we
will discuss here a single device with dimensions as indicated in
Fig.~1.
The results of the experiments are shown in Figs.~3 and 4,
corresponding to the measurement configurations of Figs.~1a and b,
respectively. In the upper three panels of the figures, the
nonlocal resistance is plotted as a function of gate 1, while in
the lower panel, gate 2 is swept.

Figure~3 corresponds to the layout of Fig.~1a, and the detected
nonlocal signal can consequently be denoted as $R_{34,12}$, i.e.
the voltage measured between contacts 1 and 2 divided by the
current passed between contacts 3 and 4. When we sweep the gate on
the injector area (gate 2) while the detector is tuned into the
QSH regime ($V^*_{\rm{gate 1}} = 0$), we observe (lower
panel in Fig.~3) a pronounced maximum around $V^*_{\rm {gate 1}} =
0$, and smaller but finite values on both sides. The signal around
$V^*_{\rm {gate 1}} = 0$ reaches approximately the quantized value
observed in our previous experiments on nonlocal transport in the
QSH regime\cite{roth2009}. We attribute the slight deviation from
perfect quantization to imperfect gating in the not gate-covered region
between gates 1 and 2. Imperfectly gated regions in the sample can
act as dephasing centers for edge electrons, which can lead to a
deviation from the expected quantized nonlocal
resistance\cite{roth2009,Maciejko2009}. In addition, in HgTe QW
devices subsequent gate voltage sweeps can charge interface trap
states in a different way\cite{Hinz2006}, leading to different
dephasing effects and a different magnitude of the deviation from
quantized resistance for each gate voltage sweep.

Apart from the large signal in the QSH regime, the measurements
also exhibit a non-vanishing nonlocal signal when the area
underneath gate 2 is metallic, either $n$- or $p$-type, and thus
corresponds to the injector region depicted in Fig.~1a. The origin
of this finite signal becomes more evident when the injector gate
voltage is set at a fixed value either in the $p$-type
($V^*_{\rm {gate 2}}=-0.75$~V~$<0$) or in the $n$-type metallic
regime ($V^*_{\rm {gate 2}}=1.0$~V~$>0$) while the voltage on gate
1 is swept (top panel of Fig.~3). Evidently, a significant
increase in the nonlocal signal is observed, with a peak when the
detector is exactly in the QSH insulator regime. This is the
observation anticipated above: one may expect a nonlocal signal
of this amplitude only when the metallic leg exhibits a SHE and
the edge channels in the QSH leg are spin polarized. Our data also
show that the nonlocal signal for the $p$-type injector
($V^*_{\rm {gate 2}}=-0.75$~V) is more than ten times larger than
that for the $n$-type injector ($V^*_{\rm {gate 2}}=1.0$~V). This
is consistent with our experimental observations on the SHE signal
in all-metallic HgTe QW\cite{Bruene2010}, where the nonlocal
signal is about an order of magnitude larger in the $p$-regime
than in the $n$-regime and results from enhanced spin-orbit
splitting in the valence band\cite{novik2005,som}.

Our data for the reverse configuration of Fig.~1b are shown in
Fig.~4. The sweep of gate 2 in the bottom panel now corresponds to
the detection leg, and one can directly see that also in this
configuration we observe a finite nonlocal signal (in this case
$R_{12,34}$), even when the detector is metallic (red and green
arrows). The upper panel shows the effect of sweeping the injector
leg (gate 1), and indicates that the nonlocal signal peaks when
the injector is in the QSH state. As in the previous configuration
of Fig.~3, we observe an order of magnitude increase in the
nonlocal signal when the metallic detector is $p$-type ($V^*_{\rm
{gate 2}}=-0.82$~V) as compared with an $n$-type detector
($V^*_{\rm {gate 2}}=1.2$~V). As noted above, our observation of
the nonlocal signal is evidence that the helical edge channels
generate a spin accumulation at the interface between the QSH
injector and the metallic detector, which responds by the
SHE$^{-1}$.

The results in Figs.~3 and 4 look very similar and, in fact, are
expected to do so on account of the Onsager-Casimir symmetry
relations for the nonlocal resistances $R_{mn,kl}$ in a four-probe
device\cite{Buttiker1986,Buttiker1988b},
\begin{align}\label{onsager}
R_{mn,kl}(\b{B})=R_{kl,mn}(-\b{B}),
\end{align}
where the first pair of indices refers to the current probes, the second pair refers to the voltage probes,
and $\b{B}$ is the
magnetic field. In the present setup, the magnetic field is zero
and we expect $R_{34,12}=R_{12,34}$. One possible explanation for
the small deviations from exact Onsager-Casimir symmetry observed
in Fig.~3 and 4 is the random charging effects of pinned
inhomogeneities (or `trap states') mentioned earlier. Two
subsequent gate voltage sweeps can result in a different interface
potential due to these charging effects\cite{Hinz2006}, which
changes the internal state of the conductor. Note however that the
symmetry between Fig.~3 and 4 is more accurate in the doped
regimes away from the nominally insulating regime, which is
expected since a higher carrier density can more effectively
screen the interface trap potentials and thus make the internal
state of the conductor less sensitive to trap charging effects.

\section{Modeling and discussion}

In order to better understand the experimental results, we have
performed semi-classical Monte Carlo calculations to obtain
a theoretical estimate of the nonlocal resistance based
on the sample geometry (Fig.~1). We focus on the setup illustrated
in Fig.~1b, where the QSH insulator acts as a spin injector and
the metallic region detects the spin polarization of edge channels
through the SHE$^{-1}$. We calculate the nonlocal resistance
$R_{12,34}$ when the current is driven between contacts $1$ and
$2$ and the voltage is measured between contacts $3$ and $4$.
$R_{12,34}$  can be expressed in terms of the transmission
coefficients\cite{Buttiker1986,Buttiker1988b} $T_{ij}$ for the
metallic region only (Eq.~S1 of the supplementary online
material). The $T_{ij}$ are calculated within the semiclassical
Monte Carlo method \cite{Beenakker1989}, which is a reasonable
approximation for Fermi wavelengths $\lambda_F\ll L$ where
$L\sim 1$~$\mu$m is the characteristic linear size of the device
(Fig.~1 and 2a). Electrons are injected at the QSH-SHE$^{-1}$ interface (yellow-green interface in Fig.~1b), and propagate quasi-ballistically into the metallic T-structure (green region in Fig.~1b) according to semiclassical equations of
motion\cite{Sundaram1999}. These equations are derived using an
effective 4-band model for HgTe QW\cite{bernevig2006d} which
explicitly contains the effects of intrinsic spin-orbit coupling due to atomic coupling between bands
\cite{som,Rothe2010}. This intrinsic spin-orbit coupling can be visualized as resulting from a Rashba field due to the edges of a typical mesa structure used in experiments \cite{Rothe2010}.
Details of the calculation are included in the supplementary
online material.

We find that the conversion of the spin signal to the electrical
signal through the SHE$^{-1}$ is dominated by the intrinsic
spin-orbit interaction while stray contributions due to voltage spreading are
negligible (see Fig.~S3 in the supplementary online material).
Fig.~5 shows the theoretical prediction of the nonlocal resistance
signal as a function of the carrier concentration in the metallic
detector. (Note that the semiclassical simulation breaks down when the chemical potential is too close to the insulating gap.) The scattering induced by the
intrinsic spin-orbit interaction is more effective when carriers
have smaller kinetic energy, and therefore smaller wave vectors at the Fermi level\cite{som}. Since the effective mass in the $p$-regime is larger than that in the $n$-regime\cite{Bruene2010}, for comparable densities the kinetic energy will be smaller in the $p$-regime. This can explain the larger nonlocal resistance signal for the $p$-regime in comparison with the $n$-regime, as well as the decrease of the signal upon increase in carrier concentration.

\section{Conclusion}

In conclusion, we have presented nonlocal transport experiments
on split-gate HgTe QW which establish for the first time the spin polarization of the helical edge states in  topological insulators. Together with our previous experiments on conductance quantization\cite{konig2007} and nonlocal edge state transport\cite{roth2009}, the present data constitute the final piece of evidence needed to fully confirm the original predictions for this novel state of matter\cite{bernevig2006d,kane2005A,bernevig2006a}. Our experiments yield a robust signal, in both experimental configurations where the helical edge channels are used for spin injection and detection. This opens a novel route for spintronics experiments in two-dimensional electron systems at zero external field.

\section{Acknowledgements}
We thank M. Leberecht and R. Rommel for assistance in some of the experiments and E. G. Novik for discussions of the band structure. We gratefully acknowledge financial support by the Deutsche Forschungsgemeinschaft (Schwerpunkt Spintronik) under the grants
     HA 5893/1-1 (EMH) and BU 1113/3-1 (HB), the German-Israeli Foundation (I-881-138.7/2005), the National Science and Engineering Research Council (NSERC) of Canada, and the Stanford Graduate Fellowship Program (SGF). SCZ is supported by the Department of Energy, Office of Basic Energy Sciences, Division of Materials Sciences and Engineering, under contract DE-AC02-76SF00515 and by the Keck Foundation. We thank the Leibniz Rechenzentrum M\"unich, the facilities of the Shared Hierarchical Academic Research Computing Network (www.sharcnet.ca) and the computing cluster of the Stanford Institute for Materials and Energy Science at the SLAC National Accelerator Laboratory for providing computer resources.

\section{Supplementary information for article `Spin polarization of the quantum spin Hall edge states'}

In this supplementary online material, we present further theoretical results and details of our semiclassical Monte Carlo calculations, as well as additional experimental data for the nonlocal resistance in structures with different injector size. These establish that the helical property of the quantum spin Hall edge states can be detected via the inverse spin Hall effect, and that the spin Hall effect can be converted into a charge signal through the helical edge channels of the quantum spin Hall state.

\section{Introduction}

This document provides the details of our theoretical approach for
the description of the hybrid quantum spin Hall -- metallic spin
Hall (QSH-SHE) `H-bar' structures illustrated in Fig.~1. of the
main text. More specifically, we calculate the nonlocal
four-terminal resistance $R_{12,34}$ in the configuration
illustrated in Fig.~1b and Fig.~\ref{fig:LB}, where the quantum
spin Hall (QSH) insulator state acts as a spin current injector,
and the metallic region as a spin current detector through the
inverse spin Hall effect (SHE$^{-1}$). The nonlocal resistance
$R_{34,12}$ in the reverse geometry (Fig.~1a), with voltage and
current probes exchanged, is the same as $R_{12,34}$ by virtue of
the Onsager-Casimir symmetry relations in the absence of a
magnetic field (Refs.~\onlinecite{Buttiker1986},
\onlinecite{Buttiker1988b} and Eq.~(1) of the main text). We find
a good qualitative agreement between the experimental and
theoretical values (Fig.~4 and 5 of the main text).

The main idea of our approach is the following. We consider the
H-bar geometry illustrated in Fig.~\ref{fig:LB}. The yellow region
on the left is in the QSH regime, while the green region on the
right is in the metallic regime, either $n$-type or $p$-type. Figure~\ref{fig:LB} represents the same setup as in Fig.~1b, where $\mathrm{L1}$ corresponds to the length of the part of the structure in the metallic regime.

We calculate the four-terminal nonlocal resistance
$R_{12,34}\equiv(V_3-V_4)/I$ where $I\equiv I_1=-I_2$ is the
current injected into lead $1$ and collected in lead $2$, and
$V_3-V_4$ is the voltage difference between leads $3$ and $4$. An
expression for $R_{12,34}$ in terms of the transmission
coefficients $T_{ij}$ from lead $j$ to lead $i$, $i,j=1,\ldots,4$,
can be derived from the multiprobe Landauer-B\"{u}ttiker
formula\cite{Buttiker1986,Buttiker1988b}, with $T_{ij}=T_{ji}$
from time-reversal symmetry\cite{Datta} (TRS). The coefficients
$T_{12}=T_{21}=1$ in the QSH region are universal and are obtained
simply by counting the number of edge states leading from one
contact to the other (see Ref. ~9). In the same spirit, the
coefficients $T_{34}=T_{43}$ can be approximated by counting the
number of channels\cite{Beenakker1991} $N_c\simeq k_FW/\pi$ (per
spin) in the vertical arm of width $W\equiv\mathrm{W2}$ in the
metallic T-structure (Fig. \ref{fig:LB}), with $k_F$ the Fermi
wave vector in the metallic region. The nontrivial part of the
calculation is to determine the coefficients
$T_{13},T_{14},T_{23}$ and $T_{24}$ for the transmission between
the contacts attached to the QSH region and those attached to the
metallic region. Since the right T-structure is metallic, for high
enough densities the Fermi wavelength $\lambda_F\ll W$ is much
smaller than the dimensions $W$ of the T-structure and a
semiclassical approach becomes justified. In other words, there is
a large number of channels $N_c\gg 1$ in the metallic region and
we can neglect the quantization of motion in the transverse
direction.

Our approach for the calculation of $T_{13},T_{14},T_{23},T_{24}$
is based on the Monte Carlo method for the simulation of
semiclassical electron transport in semiconductors\cite{Jacoboni}.
This approach has been used successfully for the study of
magnetotransport in multiprobe
conductors\cite{Beenakker1989,Beenakker1989b,Molenkamp1990,Heindrichs1998},
and in particular for the study of quasiballistic transport in
HgTe quantum wells\cite{Daumer2003}. The procedure consists in
injecting electrons with well-defined positions and momenta at the
left of the metallic T-structure; the QSH edge states then act as `injectors' of electrons into the
metallic region. The electrons then propagate into the ballistic
region according to semiclassical equations of
motion\cite{Sundaram1999}, which include elastic scattering on the
geometric boundaries of the T-structure, elastic scattering on
nonmagnetic impurities through the inclusion of a phenomenological
momentum relaxation time, and more importantly, the effect of
spin-orbit coupling through the inclusion of a Berry phase
term\cite{Sundaram1999} which acts as a magnetic field in momentum
space\cite{Sundaram1999,murakami2003,murakami2004}. This last term
is responsible for the SHE and SHE$^{-1}$. One then simply counts
the fraction of electrons which reach contacts $3$ and $4$, from
which the classical transmission probabilities
$T_{13},T_{14},T_{23},T_{24}$ can be
extracted\cite{Beenakker1989}.

This document is structured as follows. In Sec.~\ref{SecLB}, we
apply the multiprobe Landauer-B\"{u}ttiker
formula\cite{Buttiker1986} to the geometry of Fig.~\ref{fig:LB},
and give an explicit expression for the nonlocal resistance
$R_{12,34}$ in terms of calculable quantities. In
Sec.~\ref{SecEOM}, we apply the formalism of
Ref.~\onlinecite{Sundaram1999} to derive the form of the equations
of motion describing carrier propagation in the metallic region,
including the important Berry phase term. In Sec.~\ref{SecMC}, we
give the details of the Monte Carlo algorithm. In
Sec.~\ref{SecResults}, we present and discuss our numerical
results. Finally, in Sec.~\ref{AddExpTheor} we present additional experimental and theoretical results which provide strong evidence that the observed nonlocal signal is a direct consequence of spin-orbit coupling in our samples.

\section{Multiprobe Landauer-B\"{u}ttiker formalism and nonlocal resistance}
\label{SecLB}

For a four-probe phase-coherent device, the nonlocal resistance
$R_{12,34}$ is one of several possible four-terminal resistances,
whose expressions in terms of transmission coefficients were all
worked out long ago by B\"{u}ttiker\cite{Buttiker1986}. He
obtained
\begin{align}\tag{S1}\label{R1234}
R_{12,34}\equiv\frac{V_3-V_4}{I}=
\frac{h}{e^2}\frac{T_{31}T_{42}-T_{32}T_{41}}{D},
\end{align}
at zero temperature, where $I\equiv I_1=-I_2$ is the current
injected at contact $1$ and collected at contact $2$
(Fig.~\ref{fig:LB}), and
\[
D\equiv\det\left(
\begin{array}{ccc}
T_{12}+T_{13}+T_{14} & -T_{12} & -T_{13} \\
-T_{31} & -T_{32} & T_{31}+T_{32}+T_{34} \\
-T_{41} & -T_{42} & -T_{43}
\end{array}
\right),
\]
where $T_{ij}\equiv T_{i\leftarrow j}$ is the transmission
probability from lead $j$ to lead $i$ at the Fermi level, and
$T_{ij}=T_{ji}$ by TRS in the absence of a magnetic
field\cite{Datta}.

From Fig.~\ref{fig:LB}, one can read off $T_{12}=1+T'_{12}$, where
$1$ is universal and comes from the edge state propagating
directly from contact $1$ to contact $2$ along the left edge of
the QSH T-structure, while $T'_{12}$ is nonuniversal and is the
probability of an electron travelling from contact $1$ along the
top-right edge state of the QSH T-structure, entering the metallic
region, propagating inside the metallic region, returning inside
the QSH T-structure and propagating to contact $2$ via the
bottom-right edge state. In fact, $T'_{12}$ corresponds to the
probability of interedge tunneling\cite{Zhou2008}, which is
negligibly small for a wide enough device. Indeed,
Ref.~\onlinecite{Zhou2008} finds that for a device width
$\mathrm{W1}\sim 1\,\mu$m (we have $\mathrm{W1}=0.77\,\mu$m, see
Fig.~1), the gap $\Delta$ opened in the edge state dispersion by
interedge tunneling is negligibly small, $\Delta\sim 10^{-7}$ meV.
Also, nonlocal resistance measurements (similar to those of
Ref.~9 performed on the actual devices when
they are entirely gated into the QSH regime yield the values
expected from unperturbed nonlocal edge state transport (see
Ref.~9 and Fig.~2b), suggesting that interedge
tunneling is negligible. Finally, we have performed fully quantum-mechanical, numerical
calculations of the $S$-matrix of a QSH/metal interface in a strip
geometry using the tight-binding version of the four-band
effective model for the QSH state in HgTe quantum
wells\cite{Konig2008}, and confirm that $T'_{12}$ is negligible
for the sample widths considered here. Therefore we take
$T'_{12}=0$ and $T_{12}=1$.

The coefficient $T_{34}$ for transmission from contact $3$ to
contact $4$ through the metallic region (Fig.~\ref{fig:LB}) is
obtained from the semiclassical Monte Carlo calculation as
follows. We first calculate $\tilde{T}^\sigma_{34}$ and
$\tilde{R}^\sigma_3$ defined as the fraction of electrons of spin
$\sigma=\uparrow,\downarrow$ injected from contact $3$ that reach
contact $4$, or that are reflected back into contact $3$,
respectively. They satisfy
$\tilde{R}^\sigma_3+\tilde{T}^\sigma_{34}=1$ since the metal/QSH
interface is modeled as a perfectly reflecting interface and the $z$ component of the spin
is conserved in our model (the electron spin is discussed in Sec.
\ref{SecEOM} and \ref{SecMC}). However, these coefficients assume
only one transport channel (since they sum up to
one\cite{Buttiker1988b}), and furthermore neglect the probability
of being transmitted into the QSH region $T_{31}+T_{32}$. The
metallic leads have $N_c$ channels per spin at the Fermi
level\cite{Beenakker1991} with $N_c\simeq k_FW/\pi$ and
$W\equiv\mathrm{W2}$ the width of the lead (Fig.~\ref{fig:LB}).
Taking this into account as well as the probabilities
$T_{31},T_{32}$, the actual transmission and reflection
coefficients $T_{34}^\sigma$, $R_3^\sigma$ should
satisfy\cite{Beenakker1991}
$(T^\sigma_{31}+T^\sigma_{32})+T_{34}^\sigma+R_3^{\sigma}=N_c$. We
can thus construct transmission and reflection coefficients
satisfying this constraint by defining
\begin{align}
R_3^\sigma&=\tilde{R}^\sigma_3[N_c-(T_{31}^\sigma+T_{32}^\sigma)],
\tag{S2}\label{R3}
\\
T_{34}^\sigma&=\tilde{T}^\sigma_{34}[N_c-(T_{31}^\sigma+T_{32}^\sigma)].
\tag{S3}\label{T34}
\end{align}
The total transmission coefficient is then given by
$T_{34}=\sum_{\sigma=\uparrow,\downarrow}T_{34}^\sigma$.

In the Monte Carlo procedure (to be detailed in Sec.~\ref{SecMC}),
the semiclassical equations of motion, which are first order in
time, are integrated numerically, starting from the initial
positions and momenta of the charge carriers. In order to obtain
the transmission coefficients, we need to perform two separate
Monte Carlo calculations. A first calculation, where Fermi surface
electrons are injected at the QSH/metal interface and collected at
the contacts $3$ and $4$, yields $T_{13},T_{14},T_{23}$ and
$T_{24}$. The problem of the interface between a QSH
insulator and a normal metal is nontrivial and has been studied in
Ref.~\onlinecite{Yokoyama2009} using quantum-mechanical scattering
theory. In the present work, we use a simpler semiclassical approach. We take for initial conditions at the left of the metallic region a spatially uniform distribution, a fixed wave vector amplitude
$k_F=|\b{k}|$ equal to the Fermi wave vector, and an angular distribution given by
\cite{Beenakker1989,Beenakker1991}
\[
P(\theta)=\half\cos\theta,\,-\frac{\pi}{2}\leq\theta\leq\frac{\pi}{2},
\]
where $\theta$ is the angle between the carrier wave vector
$\b{k}$ and the the positive $x$ axis (Fig.~\ref{fig:LB}), and
$\int_{-\pi/2}^{\pi/2}d\theta\,P(\theta)=1$. The reason we choose a spatially uniform distribution is that although the electrons are injected from the QSH side with wave functions localized along the edge, as soon as they enter the metallic region their wave functions merge into the bulk and their localization length diverges\cite{Konig2008}. A second
calculation consists in injecting electrons from contact $3$ and
collecting them at contacts $4$ or $3$, and yields $T_{34}$
according to Eq. (\ref{T34}). The angular distribution is again given by $P(\theta)=\half\cos\theta$, $-\frac{\pi}{2}\leq\theta\leq\frac{\pi}{2}$, but now $\theta$ is the angle between the carrier wave vector
$\b{k}$ and the the negative $y$ axis (Fig.~\ref{fig:LB}). Positionwise, the
carriers are again injected uniformly across the width $\mathrm{W2}$ of
contact $3$ (Fig.~\ref{fig:LB}).

\section{Semiclassical equations of motion and Berry phase}
\label{SecEOM}

In this section we derive the equations of motion which contain
the reciprocal magnetic field in momentum space, or Berry phase
term\cite{Sundaram1999}. We neglect the bulk inversion asymmetry
terms which are a small perturbation\cite{Konig2008}. We also
neglect interband transitions and study separately the
semiclassical dynamics of Kramers partners at the Fermi level in
the metallic regime. For either of the $n$- or $p$-type regimes we
have two degenerate bands that are related by TRS, which means
that in our simulation we track the position and momentum of two
species of particles denoted by $\u$ and $\d$. The energy of the
degenerate conduction bands ($n$-type) is $E_+$ and that of the
degenerate valence bands ($p$-type) is $E_-$.

As before, we use the convention of Ref.~\onlinecite{Zhou2008} for
the Hamiltonian. The spectrum consists of two energy eigenvalues,
\begin{align}\tag{S4}\label{spectrum}
E_\pm(k^2)=\epsilon(k^2)\pm d(k^2),
\end{align}
where $\epsilon(k^2)=-Dk^2$, $d(k^2)=\sqrt{A^2k^2+M^2(k^2)}$, and
$M^2(k^2)=M-Bk^2$. Each eigenvalue is two-fold degenerate, with
eigenstates
\begin{align}
|u^\pm_\u(\b{k})\rangle&=\left(u^\pm_\u(\b{k})\right)_1|\half\rangle+
\left(u^\pm_\u(\b{k})\right)_2|\thf\rangle,\nonumber\\
|u^\pm_\d(\b{k})\rangle&=\hat{T}|u^\pm_\u(\b{k})\rangle
=\left(u^\pm_\u(\b{k})\right)_1^*|-\half\rangle+
\left(u^\pm_\u(\b{k})\right)_2^*|-\thf\rangle.\nonumber
\end{align}
The eigenspinors are given by
\[
u^\pm_\u(\b{k})\equiv
\left(
\begin{array}{c}
\left(u^\pm_\u(\b{k})\right)_1 \\
\left(u^\pm_\u(\b{k})\right)_2
\end{array}
\right)
=
\frac{1}{\sqrt{A^2k^2+g_\pm^2(k^2)}}
\left(
\begin{array}{c}
\pm A(k_x-ik_y) \\
g_\pm(k^2)
\end{array}
\right),
\]
and are orthonormal, where $g_\pm(k^2)\equiv d(k^2)\mp M(k^2)$. We
can now calculate the Berry curvatures for each
band\cite{Sundaram1999},
\begin{align}\tag{S5}\label{berry}
\Omega_{\alpha\beta}^{\pm,\sigma}(\b{k})=-2\Im\biggl\langle
\frac{\partial u_\sigma^\pm(\b{k})}{\partial k_\alpha}\biggl|
\frac{\partial u_\sigma^\pm(\b{k})}{\partial
k_\beta}\biggr\rangle,
\end{align}
with $\sigma=\uparrow,\downarrow$ and $\alpha,\beta=x,y$. We
however immediately observe that due to TRS, we have
\[
\Omega_{\alpha\beta}^{\pm,\d}(\b{k})=-\Omega_{\alpha\beta}^{\pm,\u}(\b{k}),
\]
therefore we only need to calculate the Berry curvature for spin
$\u$. Furthermore, $\Omega_{\beta\alpha}=-\Omega_{\alpha\beta}$ is
antisymmetric from the definition Eq.~(\ref{berry}). In two
dimensions, this antisymmetric tensor has a single component
$\Omega_{xy}$ and we can define a pseudoscalar
\[
\Omega^\pm(\b{k})\equiv\half\epsilon^{\alpha\beta\gamma}\Omega^{\pm,\u}
_{\beta\gamma}(\b{k})=\Omega^{\pm,\u}
_{xy}(\b{k}),
\]
where $\epsilon^{\alpha\beta\gamma}$ is the Levi-Civit\`{a} symbol
and $\alpha=z$ necessarily. We obtain
\[
\Omega^\pm(\b{k})=\Omega^\pm(k^2)=-\frac{A^2}{2}
\frac{M+Bk^2}{d(k^2)}
\frac{A^2k^2\mp 2g_\pm(k^2)M(k^2)}
{[A^2k^2\mp g_\pm(k^2)M(k^2)]^2}.
\]
Note that the physical units of the quantities $A,k,\Omega$ are
given by $[A]=\mathrm{eV}\cdot$\AA~and $[k]=$\AA$^{-1}$, hence
$[\Omega]=$\AA$^2$. The anomalous or Hall velocities are given
by\cite{Sundaram1999}
\begin{eqnarray*}
\hbar\delta\dot x_\sigma^\pm&=&-\hbar\Omega_{xy}^{\pm,\sigma}\dot k_{y,\sigma}^\pm=-\sigma\hbar\Omega^\pm\dot k_{y,\sigma}^\pm, \\
\hbar\delta\dot y_\sigma^\pm&=&-\hbar\Omega_{yx}^{\pm,\sigma}\dot k_{x,\sigma}^\pm=\sigma\hbar\Omega^\pm\dot k_{x,\sigma}^\pm,
\end{eqnarray*}
with $\sigma=\pm 1$. Since the bands are doubly degenerate, the
normal velocity is independent of spin and is simply given by
\[
\hbar\b{v}_\pm\equiv\alpha_\pm(k^2)\b{k}\text{ with }
\alpha_\pm(k^2)\equiv-2D\pm\frac{A^2-2BM(k^2)}{d(k^2)},
\]
where $[\alpha]=\mathrm{eV}\cdot$\AA$^2$. The semiclassical
equations of motion finally take the form
\begin{align}
\hbar\dot x_\sigma^\pm&=\alpha_\pm((\b{k}_\sigma^\pm)^2)k_{x,\sigma}^\pm
-\sigma\hbar\Omega^\pm((\b{k}_\sigma^\pm)^2)\dot k_{y,\sigma}^\pm,\tag{S6}\label{EOM1} \\
\hbar\dot y_\sigma^\pm&=\alpha_\pm((\b{k}_\sigma^\pm)^2)k_{y,\sigma}^\pm
+\sigma\hbar\Omega^\pm((\b{k}_\sigma^\pm)^2)\dot k_{x,\sigma}^\pm,\tag{S7}\label{EOM2} \\
\hbar\dot k_{x,\sigma}^\pm&=F^\mathrm{coll}_x
(\b{r}_\sigma^\pm,\b{k}_\sigma^\pm,t),\tag{S8}\label{EOM3} \\
\hbar\dot k_{y,\sigma}^\pm&=F^\mathrm{coll}_y
(\b{r}_\sigma^\pm,\b{k}_\sigma^\pm,t),\tag{S9}\label{EOM4}
\end{align}
where $[\hbar\dot{\b{r}}]=\mathrm{eV}\cdot$\AA, and
$\b{F}^\mathrm{coll}$ is the force exerted on the particles due to
collisions with the geometric boundaries of the sample and with
impurities in the sample. The specific form of this term is
detailed in Sec.~\ref{SecMC}.

\section{Semiclassical Monte Carlo algorithm}
\label{SecMC}

For simplicity, we assume that the probability of reflecting into
a state with opposite spin is very small, so that we are
effectively simulating the semiclassical dynamics of a
two-component gas, with the two components evolving in a perfectly
independent manner.

The collision force is given by $\b{F}^\mathrm{coll}=
\b{F}^{\partial S}+\b{F}^\mathrm{imp}$ where $\b{F}^{\partial S}$
is the force due to collision on the sample boundary (denoted by
$\partial S$), and $\b{F}^\mathrm{imp}$ is the force due to
collisions on impurities. The effect of $\b{F}^{\partial S}$ on a
particle is implemented into the simulation as follows: by energy
and momentum conservation, when a particle hits a boundary we
simply flip the sign of the component of its momentum normal to
the boundary (specular reflection). On the other hand, to take the
effect of $\b{F}^\mathrm{imp}$ into account we proceed as follows.
We first generate a random free flight time\cite{Jacoboni}
$t_\mathrm{free}$ from an exponential distribution
\[
P(t_\mathrm{free})=\frac{1}{\tau}e^{-t_\mathrm{free}/\tau},
\]
where $\tau$ is a phenomenological collision time. We expect the metallic region to be in the quasi-ballistic regime\cite{Bruene2010} and thus consider that collisions are dominated by boundary scattering. Therefore, we choose $\tau>\tau_{\partial S}$ where $\tau_{\partial S}\sim L/v_F$ is the boundary scattering time with $L$ the
characteristic linear size of the device and $v_F$ the Fermi velocity. The impurity-free equations of motion, i.e. Eqs
(\ref{EOM1})-(\ref{EOM4}) with $\b{F}^\mathrm{imp}=0$, are then
solved numerically for a time $t_\mathrm{free}$. At the end of
the free flight time, we randomize the momentum of the particle
(only the direction $\hat{k}=\b{k}/|\b{k}|$ as the magnitude
$|\b{k}|=k_F$ is fixed by energy conservation) according to an
uniform angular distribution between $0$ and $2\pi$. This is meant
to simulate isotropic scattering from rotationally invariant
impurities. Finally, a new random free flight time is generated,
and the procedure starts again until all particles have exited the
device through either contact $3$ or $4$.

Since $\hbar\dot{\b{k}}_\sigma^\pm=0$ apart from boundary and
impurity scattering, the Berry phase term vanishes for free
propagation inside the boundaries and the semiclassical
trajectories between collisions are simply straight lines. The
Berry phase term generates a shift of position upon scattering on
boundaries and impurities, an effect similar to the side-jump
effect\cite{Sinitsyn2008} in the anomalous Hall effect, with the
exception that here the spin-orbit coupling is intrinsic (arising
from the bandstructure) and not the spin-orbit coupling arising
from impurity potentials. A scattering event resulting in a change
of momentum $\Delta\b{k}$ produces a position shift $\Delta\b{r}$
given by
\begin{eqnarray*}
\Delta x_\sigma^\pm&=&-\sigma\Omega^{\pm}((\b{k}_\sigma^\pm)^2)
\Delta k_{y,\sigma}^\pm, \\
\Delta y_\sigma^\pm&=&\sigma\Omega^{\pm}((\b{k}_\sigma^\pm)^2)
\Delta k_{x,\sigma}^\pm,
\end{eqnarray*}
which can also be written
\begin{align}\tag{S10}\label{berrylorentz}
\Delta\b{r}_\sigma^\pm=-\sigma\boldsymbol{\Omega}^\pm
((\b{k}_\sigma^\pm)^2)\times\Delta\b{k}_\sigma^\pm,
\end{align}
where $\boldsymbol{\Omega}^\pm\equiv\Omega^\pm\hat{\b{z}}$ and
$\Delta \b{k}_\sigma^\pm$ depends on the boundary.
Equation~(\ref{berrylorentz}) makes explicit the interpretation of
the Berry curvature $\boldsymbol{\Omega}^\pm$ as a magnetic field
in momentum space. Since $\Delta \b{k}_\sigma^\pm$ is normal to
the boundary for specular reflection, the position shift
$\Delta\b{r}_\sigma^\pm$ will be along the tangent to the
boundary. For free propagation inside the boundaries, we have
\[
\Delta\b{r}_\sigma^\pm=\alpha_\pm((\b{k}_\sigma^\pm)^2)
\b{k}_\sigma^\pm\frac{\Delta t}{\hbar},
\]
where $[\frac{\Delta t}{\hbar}]=$~eV$^{-1}$, and $\Delta t$ is the time between two collision events (impurity or boundary). Furthermore, since we are doing a Monte Carlo simulation we need to average over a large number of particles. We find that we get reasonably good statistics (error bars not too large) for~$\sim10^6$ particles.

\section{Numerical results}
\label{SecResults}

We have performed the calculations for HgTe/HgCdTe quantum
wells\cite{Konig2008} of thickness $d=89.9$~\AA~and device size
(see Fig.~\ref{fig:LB})
$\mathrm{L1}=0.2~\mu$m, $\mathrm{W1}=0.77~\mu$m, $\mathrm{W2}=0.8~\mu$m and
$\mathrm{L2}=1.665$~\AA~(such that the distance between contacts
$3$ and $4$ is $4.1~\mu$m). These dimensions correspond to those of the device discussed in the main text (Fig.~1). Note that although the length of the middle segment of the H-bar is $0.7~\mu$m (Fig.~1), the actual length $\mathrm{L1}$ of the QSH injector region (yellow region in Fig.~1) is estimated as $0.2~\mu$m. To avoid
unphysical geometric resonances\cite{Beenakker1989,
Molenkamp1990}, we consider that the corners of the
T-structure are rounded, with radius of curvature $R=100$~\AA. For the devices studied in this work, the Dirac mass $M$ (see Sec.~\ref{SecEOM}) is estimated as $|M|=-6~$meV due to the small gap between the $H1$ and $H2$ subbands which are the lowest energy subbands for the quantum well thicknesses considered.

The results are plotted in Fig.~5 of the main text. The calculated
nonlocal signal is in good qualitative agreement with the
experimental results of Fig.~3 and 4, with $R_{12,34}\sim
10^2\,\Omega$ in the $p$-type regime and $R_{12,34}\sim
10\,\Omega$ in the $n$-type regime. Since we are performing a semiclassical
simulation for the detector region, we cannot simulate the
transition through the insulating gap, since as the Fermi level
approaches the gap, the density reaches a point where our
semiclassical approximation $k_FW/\pi\gg 1$ breaks down. Although the qualitative agreement with experiment is good, we observe that the experimental signal is larger. We expect that the discrepancy is due to the additional contribution from Rashba spin-orbit coupling\cite{Rothe2010}, which is not taken into account in the simple semiclassical approach with $S_z$ conservation used in this work.

The increase of the nonlocal resistance signal with decreasing density (Fig.~5) is evidence that the signal is generated by the spin-orbit interaction, i.e. the Berry curvature term (Sec.~\ref{SecEOM}). This can be understood if one remembers that the Berry
curvature can be treated approximately as the magnetic field in reciprocal space of a magnetic monopole\cite{murakami2003,murakami2004} centered at $\b{k}=0$. As the density is reduced (decreasing $\b{k}$), the particles are closer to the monopole and feel a stronger Berry magnetic field. Furthermore, due to the larger effective mass for holes ($p$-type) than for electrons ($n$-type) in our structures\cite{Bruene2010}, for comparable densities the Fermi wave vector will be smaller for holes than for electrons, yielding a larger Berry phase effect in the $p$-regime as compared to the $n$-regime.

\section{Additional experimental and theoretical evidence for spin-orbit origin of observed effects}\label{AddExpTheor}

We now present additional experimental and theoretical evidence that the observed nonlocal resistance signal is due to the spin-orbit interaction and not by some spurious effects. In Fig.~\ref{fig:injector_vary}, we plot the nonlocal resistance as a function of gate voltage for shorter (Fig.~\ref{fig:injector_vary}a, $200$~nm)
and longer (Fig.~\ref{fig:injector_vary}b, $400$~nm) injector sizes. The sample parameters are the same as in the main text. Although the data in Fig.~\ref{fig:injector_vary} corresponds to a metallic ($p$-type) injector, for $V^*_\mathrm{gate}\sim 0$ in the detector (QSH insulating regime) the configuration is equivalent to that illustrated in Fig.~\ref{fig:LB}, by Onsager-Casimir reciprocity (see Eq.~(1) in the main text). In our calculation, therefore, we can compare theoretical results for $L1=200$~nm and $L1=400$~nm with the data in Fig.~\ref{fig:injector_vary}a) and b), respectively.

 One can see in Fig.~\ref{fig:injector_vary} that the nonlocal resistance is essentially independent of the injector length. The maximum signal is obtained when the detector is in the QSH insulator regime, as expected from previous nonlocal transport measurements in the QSH regime (see Ref.~[9]). In the diffusive regime, we expect that a nonlocal resistance signal originating from spin effects would depend strongly on the system size. By a solution of the Poisson equation in the metallic T-structure we estimate that the stray diffusive signal for an Ohmic conductor doubles when the metallic leg of the injector is reduced by half. Therefore, a very weak dependence of the experimental nonlocal resistance on the injector length excludes the possibility that our result is a diffusive stray signal and constitutes evidence that our samples are in the quasi-ballistic regime.

We now provide further theoretical evidence that our signal originates solely from the presence of strong spin-orbit interaction effects, i.e. the Berry phase term (Sec.~\ref{SecEOM}). Fig.~\ref{fig:berry_noberry} shows the theoretically predicted nonlocal signal in the geometry of Fig.~\ref{fig:LB}. As mentioned previously, this configuration is equivalent to that of Fig.~\ref{fig:injector_vary} by Onsager-Casimir reciprocity. First, the nonlocal resistance exhibits almost no size dependence as observed experimentally (Fig.~\ref{fig:injector_vary}). Since we have chosen the bulk impurity scattering time $\tau$ to be larger than the boundary scattering time $\tau_{\partial S}$, the agreement between the theoretical and experimental results constitutes strong evidence that our samples are in the quasi-ballistic regime. Second, the nonlocal resistance signal is negligible in the absence of the Berry phase effect. Since the Berry phase term is a direct consequence of the intrinsic spin-orbit interaction (Sec.~\ref{SecEOM}), we conclude that the observed signal is chiefly due to the intrinsic SHE. However, as mentioned previously, we expect the Rashba spin-orbit interaction to also play an important role\cite{Rothe2010}, which could account for the discrepancy between our simple theory and the experiment in the magnitude of the signal. We believe that our supplementary experimental and theoretical results provide strong evidence that we have detected the helical nature of the QSH edge states via the SHE$^{-1}$, and used the QSH helical edge states to convert the SHE into a charge signal.


\begin{thebibliography}{10}
\expandafter\ifx\csname url\endcsname\relax
  \def\url#1{\texttt{#1}}\fi
\expandafter\ifx\csname urlprefix\endcsname\relax\def\urlprefix{URL }\fi
\providecommand{\bibinfo}[2]{#2}
\providecommand{\eprint}[2][]{\url{#2}}

\bibitem{bernevig2006d}
\bibinfo{author}{Bernevig, B.~A.}, \bibinfo{author}{Hughes, T.~L.} \&
  \bibinfo{author}{Zhang, S.~C.}
\newblock \bibinfo{title}{Quantum spin {Hall} effect and topological phase
  transition in {HgTe} quantum wells}.
\newblock \emph{\bibinfo{journal}{Science}} \textbf{\bibinfo{volume}{314}},
  \bibinfo{pages}{1757} (\bibinfo{year}{2006}).

\bibitem{konig2007}
\bibinfo{author}{K\"{o}nig, M.} \emph{et~al.}
\newblock \bibinfo{title}{Quantum spin {Hall} insulator state in {HgTe} quantum
  wells}.
\newblock \emph{\bibinfo{journal}{Science}} \textbf{\bibinfo{volume}{318}},
  \bibinfo{pages}{766} (\bibinfo{year}{2007}).

\bibitem{Konig2008}
\bibinfo{author}{{K\"{o}nig}, M.} \emph{et~al.}
\newblock \bibinfo{title}{The quantum spin {Hall} effect: theory and
  experiment}.
\newblock \emph{\bibinfo{journal}{J. Phys. Soc. Jpn}}
  \textbf{\bibinfo{volume}{77}}, \bibinfo{pages}{031007}
  (\bibinfo{year}{2008}).

\bibitem{kane2005A}
\bibinfo{author}{Kane, C.~L.} \& \bibinfo{author}{Mele, E.~J.}
\newblock \bibinfo{title}{Quantum spin {Hall} effect in graphene}.
\newblock \emph{\bibinfo{journal}{Phys. Rev. Lett.}}
  \textbf{\bibinfo{volume}{95}}, \bibinfo{pages}{226801}
  (\bibinfo{year}{2005}).

\bibitem{bernevig2006a}
\bibinfo{author}{Bernevig, B.~A.} \& \bibinfo{author}{Zhang, S.~C.}
\newblock \bibinfo{title}{Quantum spin {Hall} effect}.
\newblock \emph{\bibinfo{journal}{Phys. Rev. Lett.}}
  \textbf{\bibinfo{volume}{96}}, \bibinfo{pages}{106802}
  (\bibinfo{year}{2006}).

\bibitem{wu2006}
\bibinfo{author}{Wu, C.~J.}, \bibinfo{author}{Bernevig, B.~A.} \&
  \bibinfo{author}{Zhang, S.~C.}
\newblock \bibinfo{title}{Helical liquid and the edge of quantum spin {Hall}
  systems}.
\newblock \emph{\bibinfo{journal}{Phys. Rev. Lett.}}
  \textbf{\bibinfo{volume}{96}}, \bibinfo{pages}{106401}
  (\bibinfo{year}{2006}).

\bibitem{xu2006}
\bibinfo{author}{Xu, C.} \& \bibinfo{author}{Moore, J.}
\newblock \bibinfo{title}{Stability of the quantum spin {Hall} effect: effects
  of interactions, disorder, and $\mathbb{Z}_2$ topology}.
\newblock \emph{\bibinfo{journal}{Phys. Rev. B}} \textbf{\bibinfo{volume}{73}},
  \bibinfo{pages}{045322} (\bibinfo{year}{2006}).

\bibitem{Maciejko2009}
\bibinfo{author}{Maciejko, J.} \emph{et~al.}
\newblock \bibinfo{title}{{Kondo} effect in the helical edge liquid of the
  quantum spin {Hall} state}.
\newblock \emph{\bibinfo{journal}{Phys. Rev. Lett.}}
  \textbf{\bibinfo{volume}{102}}, \bibinfo{pages}{256803}
  (\bibinfo{year}{2009}).

\bibitem{roth2009}
\bibinfo{author}{Roth, A.} \emph{et~al.}
\newblock \bibinfo{title}{Nonlocal transport in the quantum spin {Hall} state}.
\newblock \emph{\bibinfo{journal}{Science}} \textbf{\bibinfo{volume}{325}},
  \bibinfo{pages}{294} (\bibinfo{year}{2009}).

\bibitem{buttiker2009}
\bibinfo{author}{{B\"{u}ttiker}, M.}
\newblock \bibinfo{title}{Edge-state physics without magnetic fields}.
\newblock \emph{\bibinfo{journal}{Science}} \textbf{\bibinfo{volume}{325}},
  \bibinfo{pages}{278} (\bibinfo{year}{2009}).

\bibitem{hankiewicz2004}
\bibinfo{author}{Hankiewicz, E.~M.}, \bibinfo{author}{Molenkamp, L.~W.},
  \bibinfo{author}{Jungwirth, T.} \& \bibinfo{author}{Sinova, J.}
\newblock \bibinfo{title}{Manifestation of the spin {Hall} effect through
  charge-transport in the mesoscopic regime}.
\newblock \emph{\bibinfo{journal}{Phys. Rev. B}} \textbf{\bibinfo{volume}{70}},
  \bibinfo{pages}{241301(R)} (\bibinfo{year}{2004}).

\bibitem{valenzuela2006}
\bibinfo{author}{Valenzuela, S.~O.} \& \bibinfo{author}{Tinkham, M.}
\newblock \bibinfo{title}{Direct electronic measurement of the spin {Hall}
  effect}.
\newblock \emph{\bibinfo{journal}{Nature}} \textbf{\bibinfo{volume}{442}},
  \bibinfo{pages}{176} (\bibinfo{year}{2006}).

\bibitem{Bruene2010}
\bibinfo{author}{Br\"{u}ne, C.} \emph{et~al.}
\newblock \bibinfo{title}{Evidence for the ballistic intrinsic spin {Hall}
  effect in {HgTe} nanostructures}.
\newblock \emph{\bibinfo{journal}{Nature Phys.}} \textbf{\bibinfo{volume}{6}},
  \bibinfo{pages}{448} (\bibinfo{year}{2010}).

\bibitem{Rothe2010}
\bibinfo{author}{Rothe, D.~G.} \emph{et~al.}
\newblock \bibinfo{title}{Fingerprint of different spin-orbit terms for spin
  transport in {HgTe} quantum wells}.
\newblock \emph{\bibinfo{journal}{New J. Phys.}} \textbf{\bibinfo{volume}{12}},
  \bibinfo{pages}{065012} (\bibinfo{year}{2010}).

\bibitem{novik2005}
\bibinfo{author}{Novik, E.~G.} \emph{et~al.}
\newblock \bibinfo{title}{Band structure of semimagnetic {Hg$_{1-y}$Mn$_y$Te}
  quantum wells}.
\newblock \emph{\bibinfo{journal}{Phys. Rev. B}} \textbf{\bibinfo{volume}{72}},
  \bibinfo{pages}{035321} (\bibinfo{year}{2005}).

\bibitem{Hinz2006}
\bibinfo{author}{Hinz, J.} \emph{et~al.}
\newblock \bibinfo{title}{Gate control of the giant {Rashba} effect in {HgTe}
  quantum wells}.
\newblock \emph{\bibinfo{journal}{Semicond. Sci. Technol.}}
  \textbf{\bibinfo{volume}{21}}, \bibinfo{pages}{501} (\bibinfo{year}{2006}).

\bibitem{som}
\bibinfo{note}{See supplementary online material.}

\bibitem{Buttiker1986}
\bibinfo{author}{B\"{u}ttiker, M.}
\newblock \bibinfo{title}{Four-terminal phase-coherent conductance}.
\newblock \emph{\bibinfo{journal}{Phys. Rev. Lett.}}
  \textbf{\bibinfo{volume}{57}}, \bibinfo{pages}{1761} (\bibinfo{year}{1986}).

\bibitem{Buttiker1988b}
\bibinfo{author}{B\"{u}ttiker, M.}
\newblock \bibinfo{title}{Symmetry of electrical conduction}.
\newblock \emph{\bibinfo{journal}{IBM J. Res. Dev.}}
  \textbf{\bibinfo{volume}{32}}, \bibinfo{pages}{317} (\bibinfo{year}{1988}).

\bibitem{Beenakker1989}
\bibinfo{author}{Beenakker, C. W.~J.} \& \bibinfo{author}{van Houten, H.}
\newblock \bibinfo{title}{Billiard model of a ballistic multiprobe conductor}.
\newblock \emph{\bibinfo{journal}{Phys. Rev. Lett.}}
  \textbf{\bibinfo{volume}{63}}, \bibinfo{pages}{1857} (\bibinfo{year}{1989}).

\bibitem{Sundaram1999}
\bibinfo{author}{Sundaram, G.} \& \bibinfo{author}{Niu, Q.}
\newblock \bibinfo{title}{Wavepacket dynamics in slowly perturbed crystals:
  gradient corrections and {Berry}-phase effects}.
\newblock \emph{\bibinfo{journal}{Phys. Rev. B}} \textbf{\bibinfo{volume}{59}},
  \bibinfo{pages}{14 915} (\bibinfo{year}{1999}).

\bibitem{Datta}
\bibinfo{author}{Datta, S.}
\newblock \emph{\bibinfo{title}{Electronic Transport in Mesoscopic Systems}}
  (\bibinfo{publisher}{Cambridge University Press},
  \bibinfo{address}{Cambridge}, \bibinfo{year}{1995}).

\bibitem{Beenakker1991}
\bibinfo{author}{Beenakker, C. W.~J.} \& \bibinfo{author}{van Houten, H.}
\newblock \bibinfo{title}{Quantum transport in semiconductor nanostructures}.
\newblock \emph{\bibinfo{journal}{Solid State Phys.}}
  \textbf{\bibinfo{volume}{44}}, \bibinfo{pages}{1} (\bibinfo{year}{1991}).

\bibitem{Jacoboni}
\bibinfo{author}{Jacoboni, C.} \& \bibinfo{author}{Lugli, P.}
\newblock \emph{\bibinfo{title}{The Monte Carlo Method for Semiconductor Device
  Simulation}} (\bibinfo{publisher}{Springer Verlag}, \bibinfo{address}{New
  York}, \bibinfo{year}{1989}).

\bibitem{Beenakker1989b}
\bibinfo{author}{Beenakker, C. W.~J.} \& \bibinfo{author}{van Houten, H.}
\newblock \bibinfo{title}{Magnetotransport and nonadditivity of point-contact
  resistances in series}.
\newblock \emph{\bibinfo{journal}{Phys. Rev. B}} \textbf{\bibinfo{volume}{39}},
  \bibinfo{pages}{10 445} (\bibinfo{year}{1989}).

\bibitem{Molenkamp1990}
\bibinfo{author}{Molenkamp, L.~W.} \emph{et~al.}
\newblock \bibinfo{title}{Electron-beam collimation with a quantum point
  contact}.
\newblock \emph{\bibinfo{journal}{Phys. Rev. B}} \textbf{\bibinfo{volume}{41}},
  \bibinfo{pages}{1274} (\bibinfo{year}{1990}).

\bibitem{Heindrichs1998}
\bibinfo{author}{Heindrichs, A.}, \bibinfo{author}{Buhmann, H.},
  \bibinfo{author}{Godijn, S.~F.} \& \bibinfo{author}{Molenkamp, L.~W.}
\newblock \bibinfo{title}{Classical rebound trajectories in nonlocal ballistic
  electron transport}.
\newblock \emph{\bibinfo{journal}{Phys. Rev. B}} \textbf{\bibinfo{volume}{57}},
  \bibinfo{pages}{3961} (\bibinfo{year}{1998}).

\bibitem{Daumer2003}
\bibinfo{author}{Daumer, V.} \emph{et~al.}
\newblock \bibinfo{title}{Quasiballistic transport in {HgTe} quantum-well
  nanostructures}.
\newblock \emph{\bibinfo{journal}{Appl. Phys. Lett.}}
  \textbf{\bibinfo{volume}{83}}, \bibinfo{pages}{1376} (\bibinfo{year}{2003}).

\bibitem{murakami2003}
\bibinfo{author}{Murakami, S.}, \bibinfo{author}{Nagaosa, N.} \&
  \bibinfo{author}{Zhang, S.~C.}
\newblock \bibinfo{title}{Dissipationless quantum spin current at room
  temperature}.
\newblock \emph{\bibinfo{journal}{Science}} \textbf{\bibinfo{volume}{301}},
  \bibinfo{pages}{1348} (\bibinfo{year}{2003}).

\bibitem{murakami2004}
\bibinfo{author}{Murakami, S.}, \bibinfo{author}{Nagaosa, N.} \&
  \bibinfo{author}{\textrm{S.C. Zhang}}.
\newblock \emph{\bibinfo{journal}{Phys. Rev. B}} \textbf{\bibinfo{volume}{69}},
  \bibinfo{pages}{235206} (\bibinfo{year}{2004}).

\bibitem{Zhou2008}
\bibinfo{author}{Zhou, B.}, \bibinfo{author}{Lu, H.-Z.}, \bibinfo{author}{Chu,
  R.-L.}, \bibinfo{author}{Shen, S.-Q.} \& \bibinfo{author}{Niu, Q.}
\newblock \bibinfo{title}{Finite size effects on helical edge states in a
  quantum spin-{Hall} system}.
\newblock \emph{\bibinfo{journal}{Phys. Rev. Lett.}}
  \textbf{\bibinfo{volume}{101}}, \bibinfo{pages}{246807}
  (\bibinfo{year}{2008}).

\bibitem{Yokoyama2009}
\bibinfo{author}{Yokoyama, T.}, \bibinfo{author}{Tanaka, Y.} \&
  \bibinfo{author}{Nagaosa, N.}
\newblock \bibinfo{title}{Giant spin rotation in the junction between a normal
  metal and a quantum spin {Hall} system}.
\newblock \emph{\bibinfo{journal}{Phys. Rev. Lett.}}
  \textbf{\bibinfo{volume}{102}}, \bibinfo{pages}{166801}.

\bibitem{Sinitsyn2008}
\bibinfo{author}{Sinitsyn, N.~A.}
\newblock \bibinfo{title}{Semiclassical theories of the anomalous {Hall}
  effect}.
\newblock \emph{\bibinfo{journal}{J. Phys. Condens. Matter}}
  \textbf{\bibinfo{volume}{20}}, \bibinfo{pages}{023201}
  (\bibinfo{year}{2008}).

\end{thebibliography}

\begin{figure}[tbh]
\includegraphics[width=5.5in]{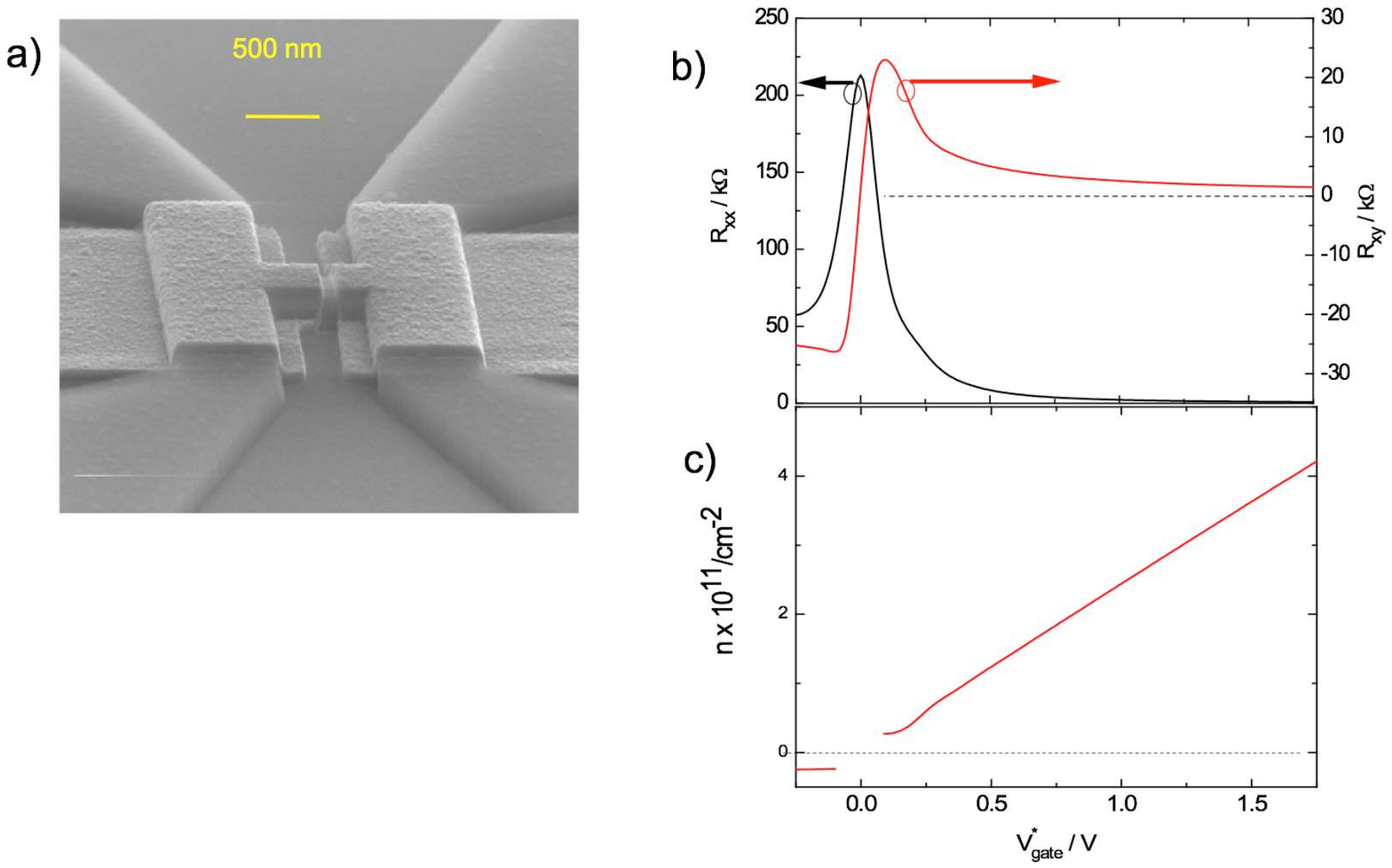}
\caption{(a) An electron micrograph of the actual device structure (rotated by 90 degrees compared to Fig. 1).
(b) and (c)  Gate voltage dependence of the longitudinal resistance R$_{\rm xx}$ (black) and Hall resistance R$_{\rm xy}$ (red) at B = 1 T, and the inferred carrier density, n, of a macroscopic Hall bar, 600 $\mu$m x 200 $\mu$m in size, fabricated from the same HgTe wafer as the nanostructures used in the experiments of Figs. 3 and 4.}
\end{figure}

\begin{figure}[tbh]
\includegraphics[width=5.5in]{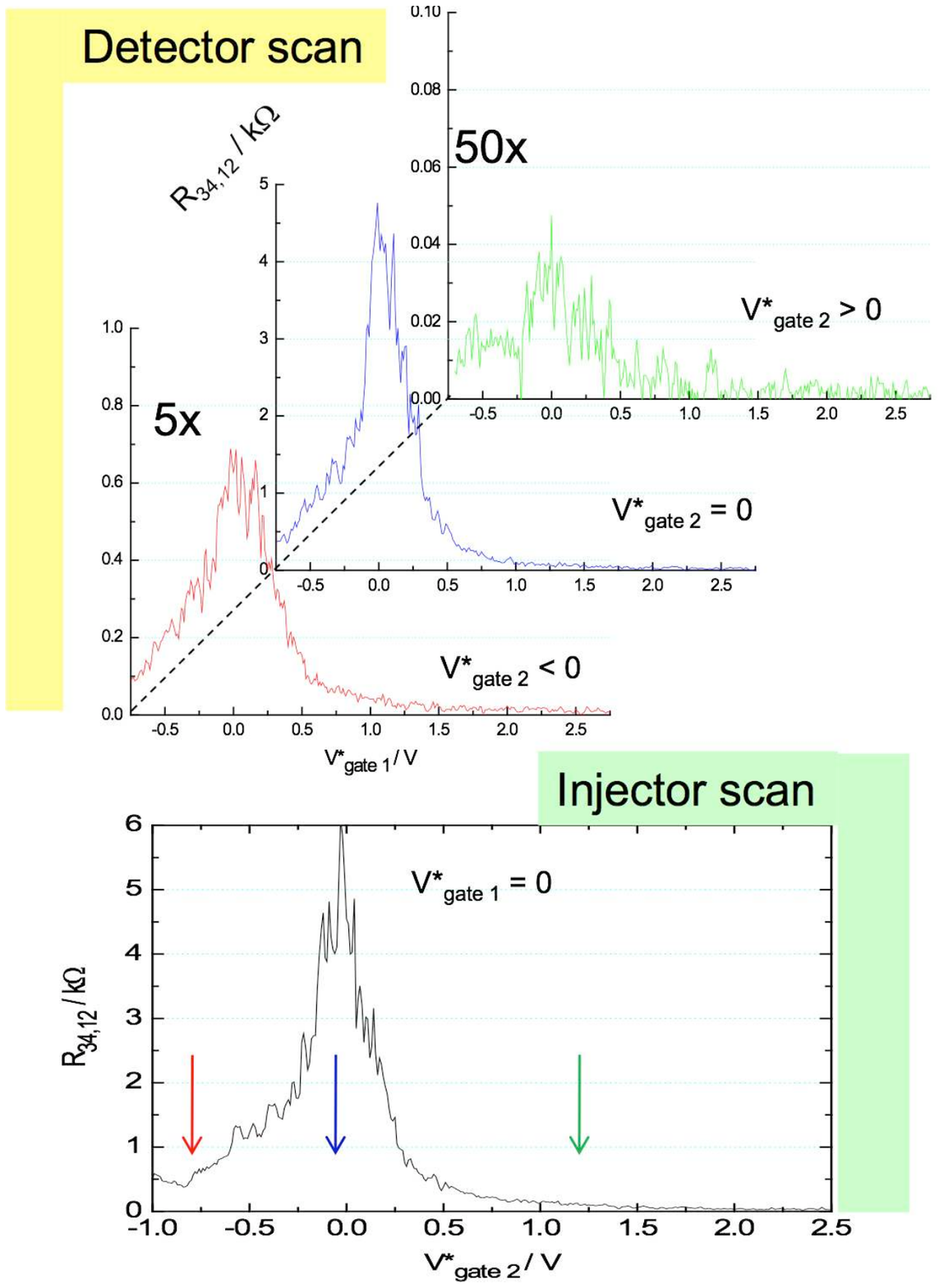}
\caption{Experimental nonlocal resistance data corresponding
to the measurement configuration of Fig.~1 a). In the bottom (green) panel, the gate on the
current injection leg is swept, varying the area from $p$- to $n$-metallic conductance, while the detector (top) leg is kept in the middle of the QSH insulator regime. The red, blue and green arrows denote gate voltages where the injector region is $p$-type metallic, QSH insulating and $n$-type metallic, respectively. In the top panel, the gate in the detector area is varied at exactly these injector settings.}
\end{figure}

\begin{figure}[tbh]
\includegraphics[width=5.5in]{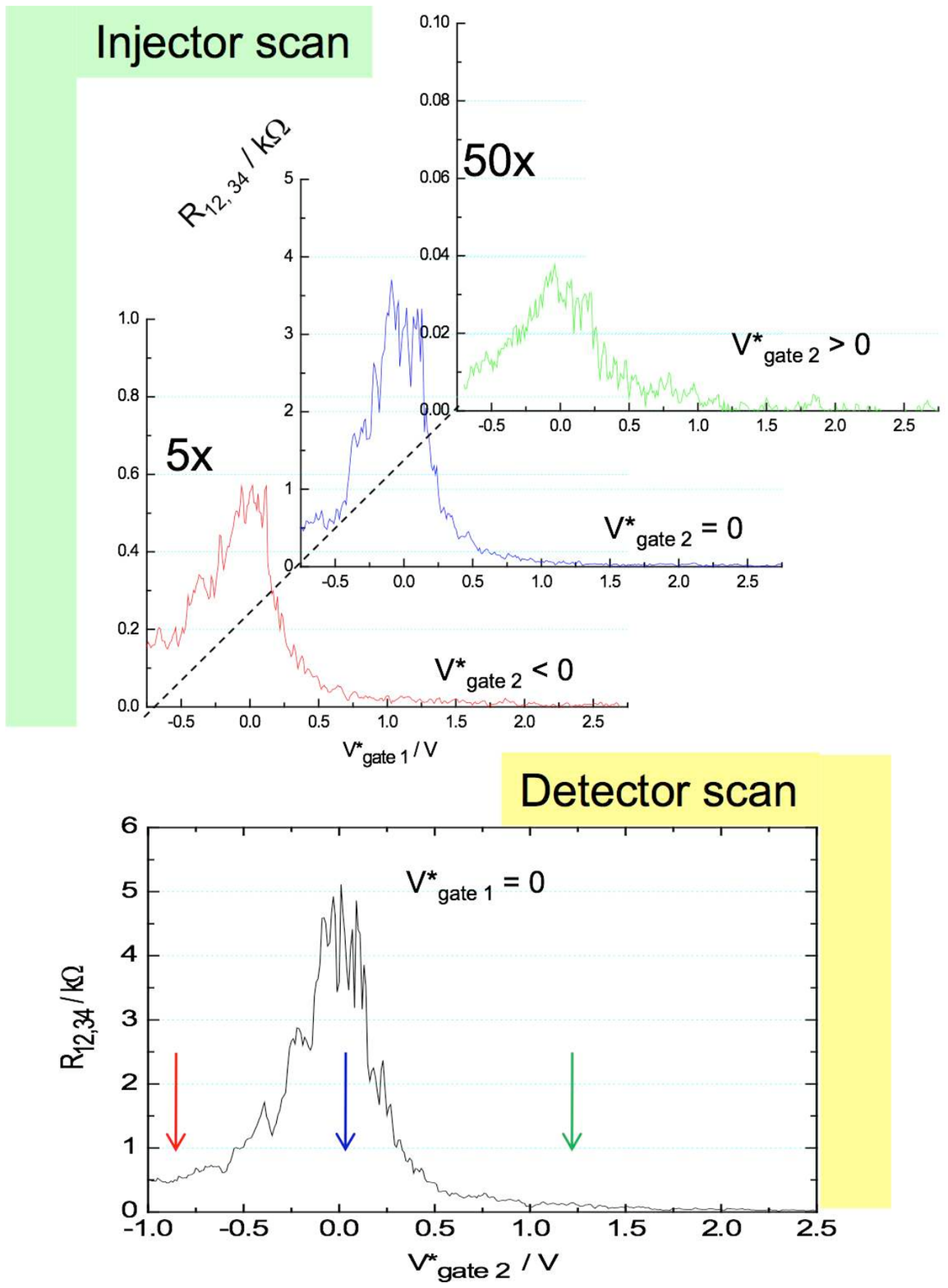}
\caption{Experimental nonlocal resistance data corresponding
to the measurement configuration of Fig.~1 b).
In the bottom (yellow) panel, the gate on the
detection leg is swept, varying the area from $p$- to $n$-metallic conductance, while the injector (bottom) leg is kept in the middle of the QSH insulator regime. The red, blue and green arrows denote gate voltages where the detector region is $p$-type metallic, QSH insulating and $n$-type metallic, respectively. In the top panel, the gate in the injector area is varied at exactly these detector settings. }
    \end{figure}

\begin{figure}[tbh]
\includegraphics[width=5.5in]{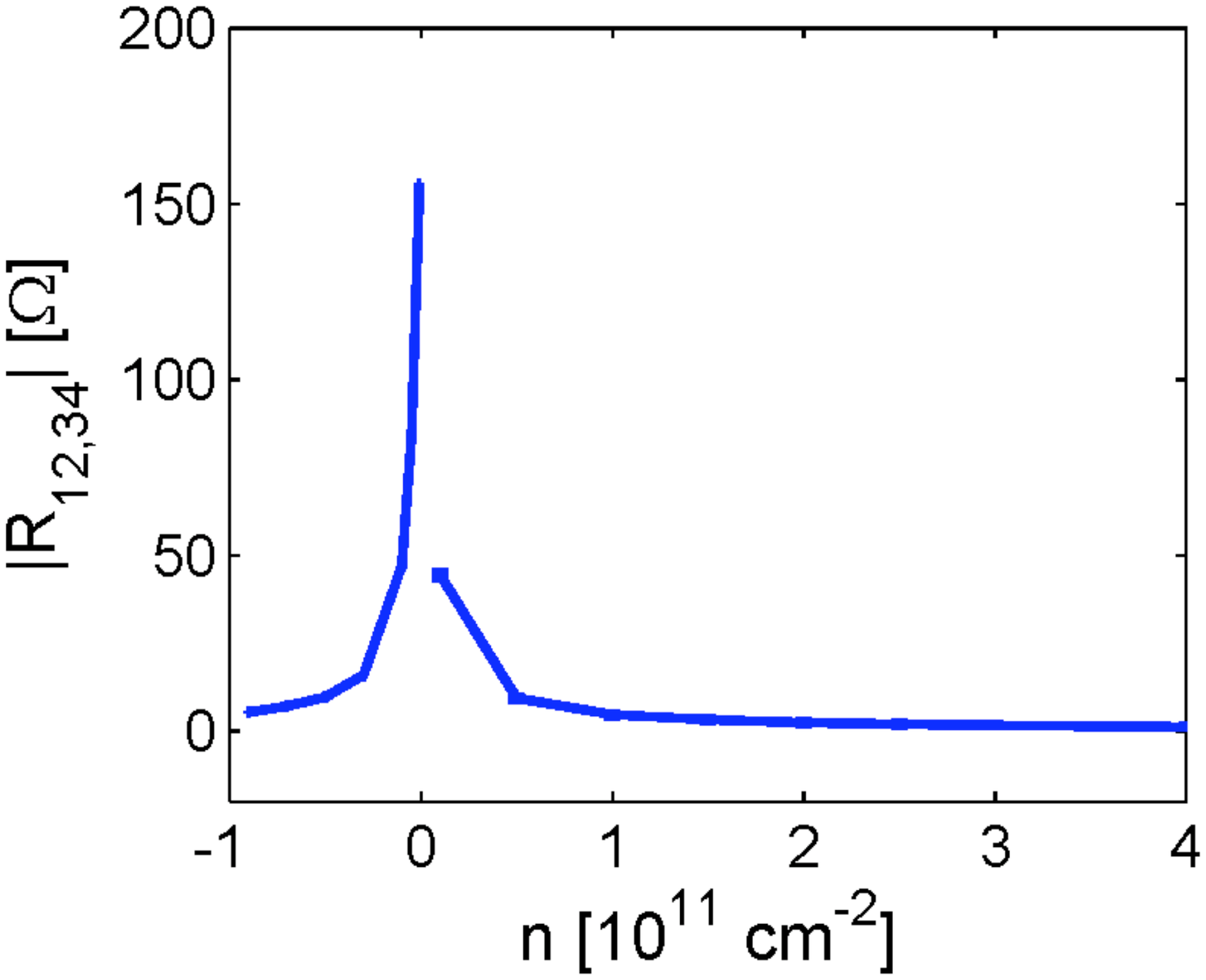}
\caption{Semiclassical Monte Carlo simulation of the nonlocal resistance signal in the setup of Fig.~1b, as a function of carrier concentration in the metallic detector.}
\end{figure}

\renewcommand{\thefigure}{S\arabic{figure}}
\begin{figure}[h]
\begin{center}
\includegraphics[width=6in]{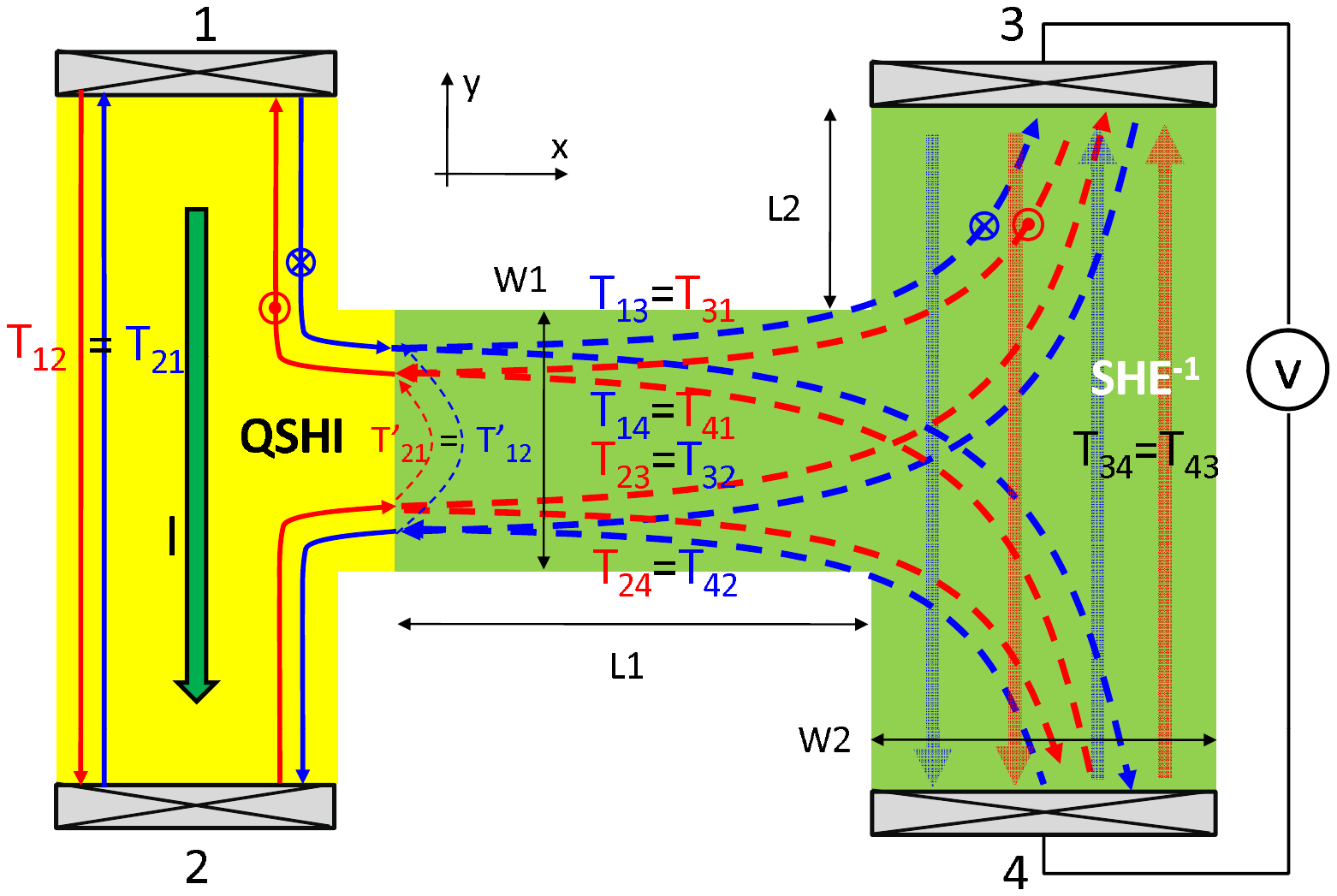}
\end{center}
\caption{Transmission coefficients for Landauer-B\"{u}ttiker calculation
of nonlocal resistance $R_{12,34}$, in the geometry corresponding
to the QSH state as spin current injector, with the metallic state exhibiting
the SHE$^{-1}$ as a spin current detector (similar to
Fig.~1b of main text).}
\label{fig:LB}
\end{figure}

\begin{figure}[tbh]
\begin{center}
\includegraphics[width=5in]{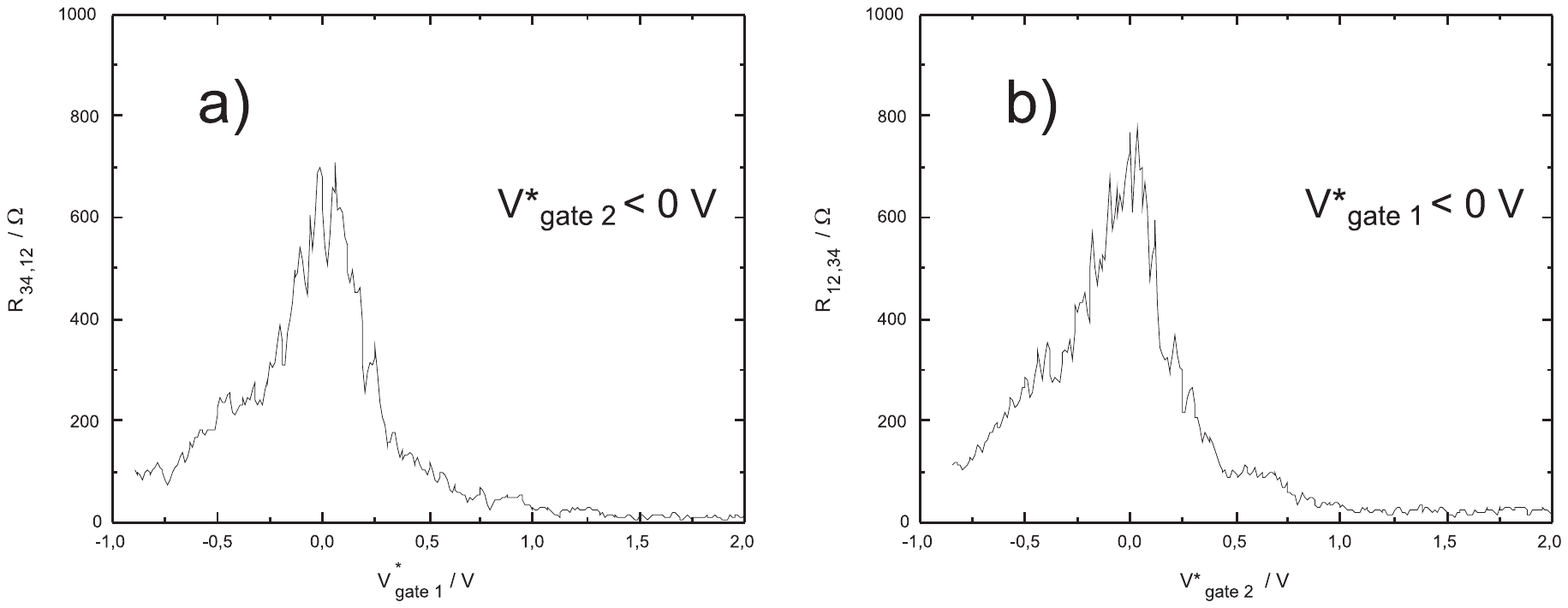}
\end{center}
\caption{ Nonlocal resistance signal for injector in the $p$-type regime, as a function of gate voltage in the detector and injector length: a) short $p$-type injector ($200$~nm); b) long $p$-type injector ($400$~nm). By Onsager-Casimir reciprocity, for the detector in the QSH insulating regime ($V^*_\mathrm{gate}\sim 0$) this is equivalent to the configuration of Fig.~\ref{fig:LB} with $L1=200$~nm in a) and $L1=400$~nm in b).}
\label{fig:injector_vary}
\end{figure}

\begin{figure}[t]
\begin{center}
\includegraphics[width=5.8in]{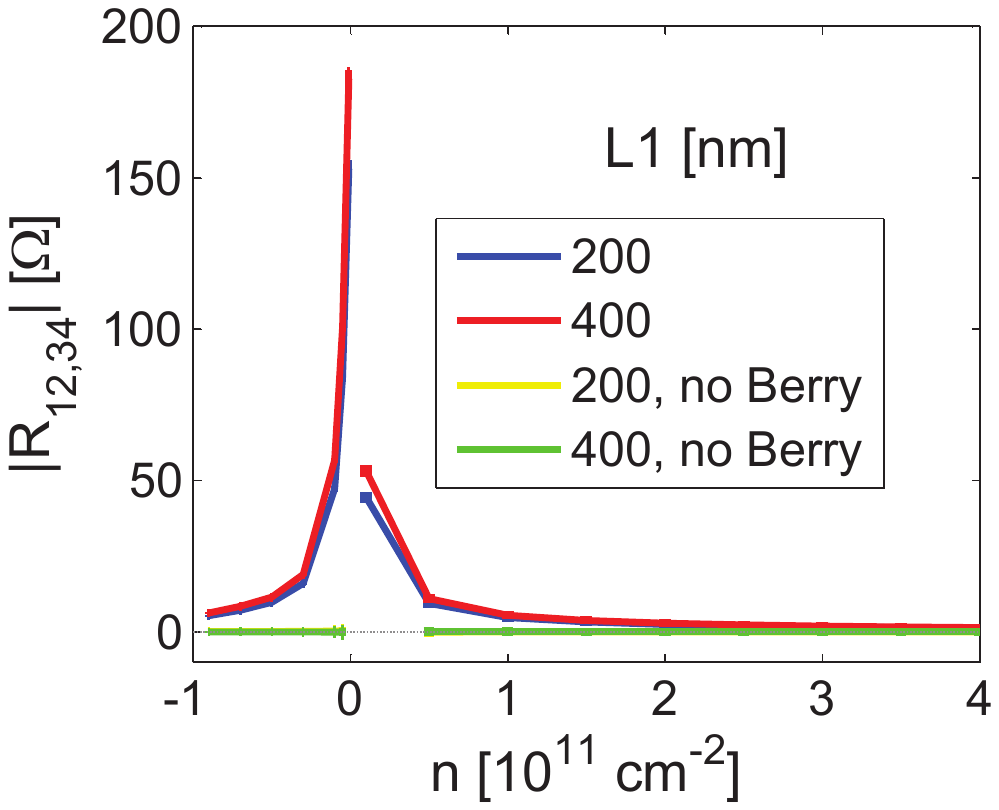}
\end{center}
\caption{Effect of the detector size $L1$ (geometry of Fig.~\ref{fig:LB}, equivalent by Onsager-Casimir reciprocity to Fig.~\ref{fig:injector_vary}) and Berry phase term on the nonlocal resistance. The dotted gray line denotes zero. The signal is essentially independent of $L1$, in agreement with experiment (Fig.~\ref{fig:injector_vary}). Furthermore, the signal is essentially zero in absence of the Berry phase term.}
\label{fig:berry_noberry}
\end{figure}
\end{document}